\documentclass[journal]{IEEEtran}
\ifCLASSINFOpdf
   \usepackage[pdftex]{graphicx}
   \graphicspath{{../pdf/}{../jpeg/}}
   \DeclareGraphicsExtensions{.pdf,.jpeg,.png}
\else
   \usepackage[dvips]{graphicx}
   \graphicspath{{../eps/}}
   \DeclareGraphicsExtensions{.eps}
\fi
\usepackage{epstopdf}
\usepackage{amsmath}
\usepackage{bm}
\usepackage{amsthm}
\UseRawInputEncoding
\usepackage{subfigure}
\usepackage{amssymb}
\usepackage{cite}
\usepackage{color}

\begin{document}
\title{Unsynchronized Reconfigurable Intelligent Surfaces with Pulse-Width-Based Design}
 \author{Ashif Aminulloh Fathnan, Kairi Takimoto, Mizuki Tanikawa, Kazutomo Nakamura, Shinya Sugiura and~Hiroki Wakatsuchi% <-this % stops a space
  \thanks{Manuscript received December 19, 2022. This work was supported in part by the Japanese Ministry of Internal Affairs and Communications (MIC) under the Strategic Information and Communications R\&D Promotion Program (SCOPE) (No. 192106007), the Japan Science and Technology Agency (JST) under the Precursory Research for Embryonic Science and Technology (PRESTO) Program (No. JPMJPR1933 and JPMJPR193A), KAKENHI grants from the Japan Society for the Promotion of Science (JSPS) (No. 21H01324 and 22F22359), and National Institute of Information and Communications Technology (NICT), Japan under the commissioned research (No. 06201).}
  \thanks{A. A. Fathnan, K. Takimoto, M. Tanikawa, K. Nakamura and H. Wakatsuchi are with the Graduate School of Engineering, Nagoya Institute of Technology, Nagoya, Aichi, 466-8555, Japan (e-mail: fathnan.aminulloh@nitech.ac.jp).}% <-this % stops a space
  \thanks{S. Sugiura is with the Institute of Industrial Science, The University of Tokyo, Tokyo 153-8505, Japan.}%
}
% The paper headers
% ====================================================================
\maketitle

\begin{abstract}
In this correspondence, we present a reconfigurable intelligent surface (RIS) that reflects an incident signal to the desired direction, depending on the pulse width. This unique reconfigurability emerges by virtue of the transient response of the embedded nonlinear circuits within the RIS unit cells. The RIS does not need any control lines connected to the unit cells or precise synchronization with the base station, leading to reduced complexity of the RIS system yet allowing automated beamforming characteristics in accordance with the incident pulse width. To verify this scheme, an RIS system model using binary phase shift keying is considered, in which the RIS wireless link is connected to a single transmitter at the input and two receivers at the output. Numerical demonstrations show that modulated signals are received with contrasting bit-error rates in the two receivers, consistent with the aforementioned pulse-width-based beam control.
\end{abstract}


\section{Introduction}

\IEEEPARstart{F}{uture} wireless networks are driving toward fulfilling various necessities, from providing higher data rates and seamless user mobility to sensing and analyzing data from the surrounding environment \cite{gui20206g}. While it may be possible to address these challenges within the boundaries of classical solutions, such as increasing the transmitter power and adding more base stations, these concepts impose more stringent requirements on energy and cost efficiency. Recently, a new concept for wireless communication systems called ‘smart radio environments’ has been proposed to address these challenges \cite{di2020smart}. This new concept is based on the reconfigurable intelligent surfaces (RISs) consisting of small, periodic, engineered wave scatterers. In this new wireless network architecture, RISs are placed in many locations within the target environment to interact favorably with radio waves, thereby improving the overall network performance.

% =======
% FIG. 01
% =======
\begin{figure}
  \begin{center}
  \includegraphics[width=0.86\linewidth]{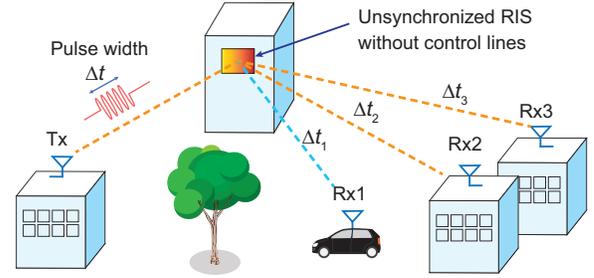}\\
  \caption{Illustration of the proposed RIS for outdoor communication. The RIS control the beamforming operation depending on the pulse width of the incident wireless signals.}\label{fig-1}
  \end{center}
\end{figure}

RISs typically utilize the concept of metasurface beamforming based on resonators with subwavelength dimensions \cite{holloway2012overview}. For realizing beamforming RISs, various practical schemes have been proposed, such as using diode-based copolarized reconfigurable unit cells 
\cite{zhang2020beyond,dai2020reconfigurable}, dynamic phase profiles with polarization conversion \cite{sugiura2021joint} or time-variable unit cells enabling frequency shifts \cite{zhang2021wireless}. While these RISs have demonstrated various wave manipulation abilities, the alteration of the RIS states typically requires a dedicated controller and complex biasing. Precise synchronization between a base station (BS) and a receiver is a challenging task, especially in a fast-changing environment with a short coherence time, due to the heavy reliance of the RIS on feedback from the receiver \cite{elmossallamy2020reconfigurable}. A small number of sensing devices with a pilot-based channel estimator may be included in the RIS system, offering a solution to the synchronization problem \cite{taha2021enabling}. However, the presence of multiple control lines between the sensors and the unit cells increases the hardware complexity and cost, especially with a large number of elements.

In view of the above challenges, we propose an RIS whose unit cells self-tune based on the incoming waveform types. Unlike typical diode-based RISs, which require biasing from an external control entity, the unit cell here is controlled directly by embedded circuits that are sensitive to the pulse width of the incoming wireless signals. Therefore, reconfigurability is obtained through passive tuning (rather than active tuning) of the unit cells. The resulting RIS performs dynamic beamforming without the need for complicated control lines and precise synchronization with the BS. This concept is based on waveform-selective metasurfaces as reported previously for absorbers, surface-wave beamformers, and Fresnel lenses \cite{wakatsuchi2013waveform,homma2022anisotropic,fathnan2022method}, among other applications. As depicted in Fig.~1, by placing these metasurfaces within a wireless environment, an RIS system may be designed that can selectively steer communication radio waves through pulse-width-based control.

% === II. RIS System and Unit-Cell Design ========================
% =========================================
\section{RIS System and Unit Cell Design}
\subsection{Passive Reconfigurable Beamforming Concept}
The RIS considered here utilizes the beamforming technique to modify the propagation of impinging radio waves through passive tuning of the unit cells. An arrangement of subwavelength reflecting elements is utilized, in which the RIS can be realized as a simple diffraction grating. In this configuration, reflective elements are periodic and are used to split an incident wave into several diffraction orders. The relationship among the angle of the diffracted beam ($\theta_m$), the incidence angle ($\theta_i$), the order of diffraction (\textit{m}), and the period of the scatterers (\textit{P}) can be written as \cite{o2004diffractive}
\begin{equation}
\theta_m= \arcsin\Big(\frac{m\lambda}{P}-\sin{\theta_{i}}\Big).
\end{equation}
As an alternative approach, an RIS can be implemented by using a gradient phase profile to realize anomalous reflection, typically maximizing the first diffraction order ($m = 1$). In such a case, all unit cells within the period of $P={\lambda}/({\mathrm{sin}\,\theta_R-\mathrm{sin}\,\theta_i})$ are set to be maximum with a phase variation of \cite{yu2011light}
\begin{equation}
 \phi(x)=\frac{2\pi{f}}{c}x\,\mathrm{sin}\,\theta_R,
\end{equation}
where $\phi(x)$ describes the variation of the gradient phase profile across unit cells located along the $x$ coordinate, $\theta_{R}$ is the reflection angle, $f$ is the frequency and $c$ is the speed of light. Note that $\theta_R$ here is the angle of the first diffraction order, i.e., $\theta_R=\theta_1$. While this anomalous reflection is optimized when all unit cells have the maximum reflection amplitude, it is possible to reconfigure some unit cells to have a low reflection amplitude (high absorption) such that the RIS becomes only sparsely radiating. In this case, the RIS switches from acting as an anomalous reflector (reflecting on one optimized angle) to acting as a diffraction grating (reflecting on several discrete angles).

\begin{figure}[t]
  \begin{center}
  \includegraphics[width=\linewidth]{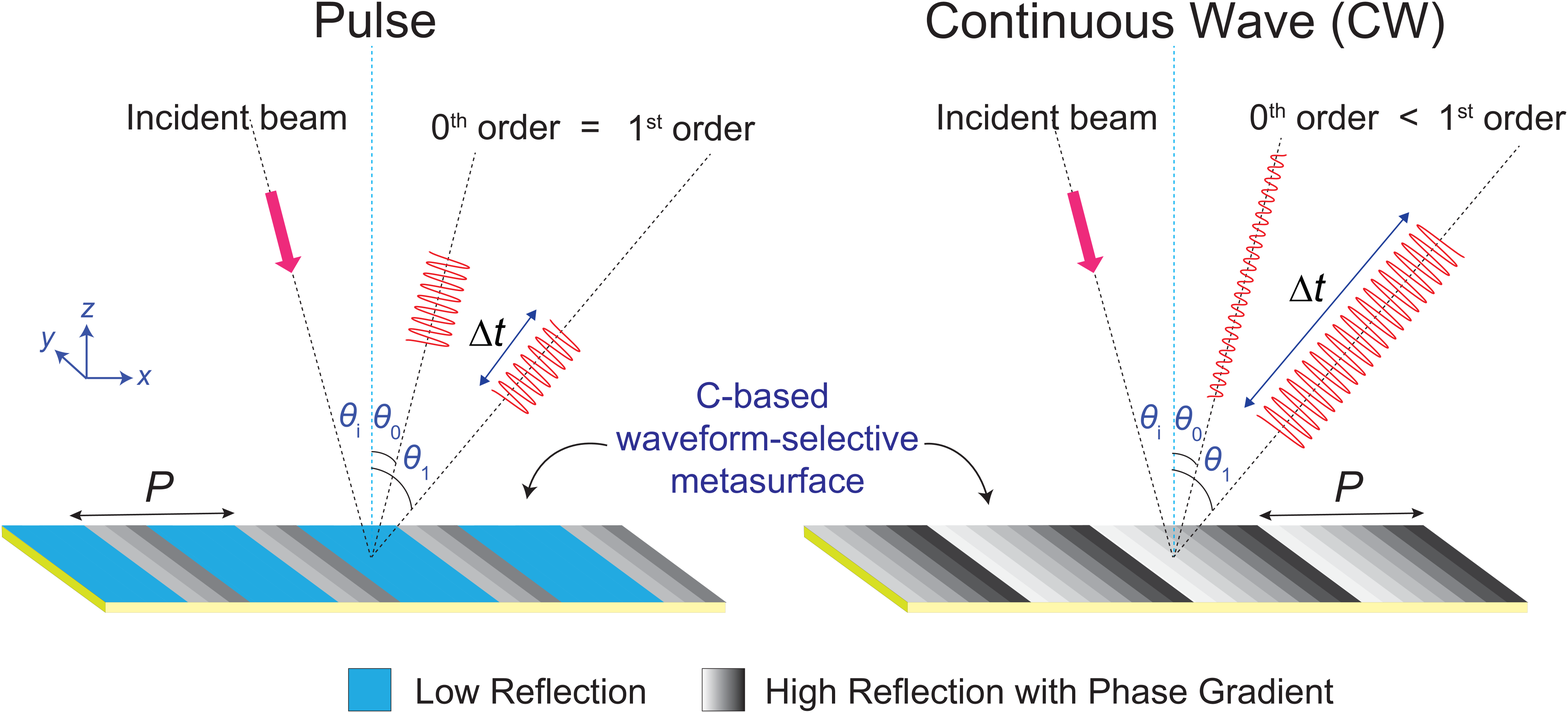}
  \caption{Passive reconfigurable beamforming concept used in the proposed RIS. An incident pulse is reflected with the same power at the zeroth and first diffraction orders, whereas a continuous incident wave is reflected mainly at the first diffraction order, while the reflection at the zeroth diffraction order is suppressed. }\label{fig-2}
  \end{center}
\end{figure}

Consistent with the above scenario, the RIS is designed to alter its beamforming operation depending on the incident pulse width. As illustrated in Fig.~2, we design a metasurface with a $2\pi$ phase gradient for a continuous incident wave (with a sufficiently long pulse width) such that the reflection at the anomalous angle (the first diffraction order) is maximized. However, upon the arrival of a short pulse ($\Delta{t}$), the unit cells are self-tuned to have low-reflection regions such that the metasurface behaves as a normal diffraction grating, resulting in equal reflected power at the zeroth and the first diffraction orders. In this scenario, the RIS passively performs unicasting for continuous wave signals and broadcasting for pulsed signals based on the incident pulse width.

\subsection{Unit Cell Design}
For our RIS design, we consider a metasurface that operates at 3\,GHz ($\lambda = 100$\,mm), with eight unit cells for one period of $P = 160$\,mm. Each unit cell has a lateral size of $L_1 = 20$\,mm. As illustrated in Fig.~2, if a continuous wave (CW) propagates with an incidence angle of $\theta_i=-10^\circ$ to the metasurface's normal plane, in accordance with (2), reflection at $\theta_1 = 53^\circ$ (the first diffraction order) will be optimized. Otherwise, when the metasurface receives short incident pulses, some unit cells will have a low reflection amplitude (high absorption). As expressed in (1), this condition creates sparsely radiating RIS unit cells that have the same reflection power at both the specular and anomalous angles of $\theta_0 = 10^\circ$ (the zeroth diffraction order) and $\theta_1 = 53^\circ$ (the first diffraction order).

\begin{figure}
  \begin{center}
  \includegraphics[width=\linewidth]{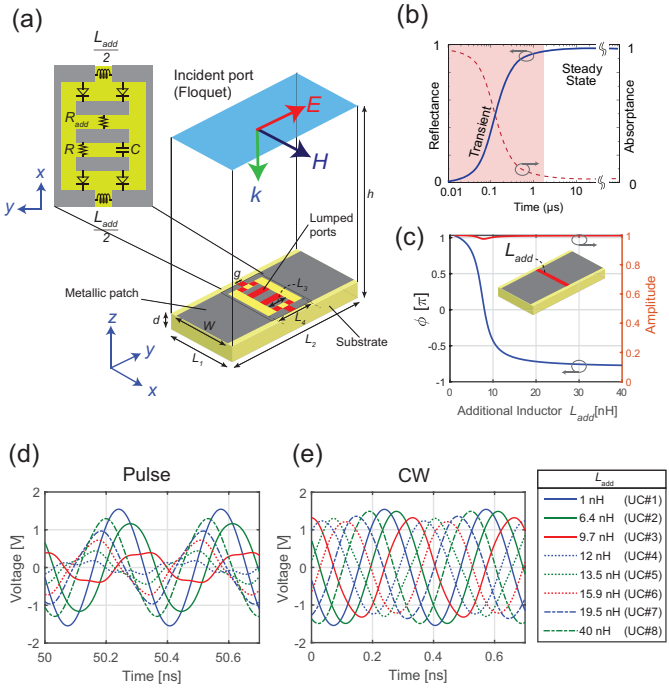}
  \caption{(a) Unit cell design of an RIS using a C-based waveform-selective metasurface, where $d =3.04$\,mm, $h = 150$\,mm, $g = 1$\,mm, $L_1 = 20$\,mm, $L_2 = 30$\,mm, $L_3 = 8$\,mm, $L_4 = 2.2$\,mm, $C = 100$\,pF, $R = 50$\,k$\Omega$, and $R_{add} = 300$\,$\Omega$. (b) Ideal profile of the C-based waveform-selective metasurface, where the reflectance increases with time and the absorptance decreases with time. (c) Phase and amplitude response of a simple metallic patch with the same geometric parameters as the adopted unit cell design. (d) Reflected voltage from the unit cell during the initial time at approximately 50\,ns (pulse response). (e) Reflected voltage from the unit cell at a much later time (CW response). \label{fig-3}}
  \end{center}
\end{figure}

Such operation can be realized by using C-based waveform-selective metasurfaces, as these metasurfaces are characterized by a pulse-width-dependent reflection amplitude \cite{wakatsuchi2019waveform}. Here, we modify the metasurface configuration presented in Ref.~\cite{wakatsuchi2019waveform} by adopting the unit cell design illustrated in Fig.~\ref{fig-3}(a). The unit cell consists of an etched metallic layer on top of a grounded dielectric substrate (Rogers RO3003, $\epsilon_r = 3$). Within the etched metallic layer, several lumped components are connected, i.e., four diodes forming a full-wave rectifier with a parallel resistive--capacitive (RC) load in series with an additional resistance $R_{add}$. The inset of Fig.~\ref{fig-3}(a) shows the detailed circuit connections. The diodes here are modeled using the SPICE model based on Avago's HSMS 286x Schottky diode series (see Ref.~\cite{wakatsuchi2019waveform}). When this circuit is embedded into the metallic patches, slow time variation of the reflectance from the metasurface is observed, as illustrated in Fig.~\ref{fig-3}(b), as a result of the transient response of the reactive RC load. In the transient regime, the reflectance of the metasurface slowly increases until it reaches the maximum level in the steady-state condition (solid line) \cite{wakatsuchi2019waveform}. Conversely, the absorption of the metasurface slowly decreases and stabilizes at the minimum level under the steady-state condition (dashed line). Therefore, any incident pulses with a short pulse width ($\Delta{t} <$ the time constant $\tau$) are reflected with a low-amplitude power, while long pulses ($\Delta{t} \gg \tau$) are reflected with a high-amplitude power.
Such a metasurface can therefore selectively reflect incident waves depending on the waveform type or the pulse width ($\Delta{t}$).

Furthermore, to obtain the gradient phase profile required to implement beamforming, the resonance frequency of the metasurface is tuned by means of additional reactive components. As seen in the inset of Fig.~\ref{fig-3}(a), we add two inductors at the corners of the top patches, each having a value of $\frac{L_{add}}{2}$. By carefully choosing $L_{add}$ to implement the phase profile in (2), we can design a metasurface that anomalously reflects an incident wave in the steady-state condition. This is similar to the way in which a simple patch structure can be tuned to realize beamforming in linear operation, i.e., without diodes or additional lump components. As seen in the inset of Fig.~3(c), using the same patch geometry, we can estimate the required $L_{add}$ for a particular beamforming operation. Here, $2\pi$ reflection phase coverage is obtained when $L_{add}$ is varied from 1 to 40\,nH.

To evaluate the unit cell response, we use the co-simulation method available in ANSYS Electronics Desktop 2020 R2. Similar to the method used in \cite{wakatsuchi2019waveform}, here, we first simulate the unit cell structure shown in Fig.~3(a) using the full-wave electromagnetic (EM) simulator HFSS. Periodic boundary conditions are used, with the incident port working as a Floquet port. After the full-wave EM simulation has run, the result is imported into a circuit simulation in ANSYS Circuit, where the lumped elements (red ports in Fig.~3(a)) are treated as open port connectors. Via these ports, the embedded circuits are connected, including diodes, the inductance $L_{add}$, two resistances $R$ and $R_{add}$, and a capacitance $C$. 

Simulation results are shown in Fig.~\ref{fig-3}(d) and (e) for eight chosen unit cells, showing the time-varying voltages under both the transient condition (short pulse, approximately 50\,ns) and the steady-state condition (very long pulse, equivalent to a CW). Note that as a result of using a harmonic balance analysis functionality, in Fig.~\ref{fig-3}(e), the time starts at 0\,ns, but it is referenced with respect to a significantly long pulse (in practice, it starts at $t \gg \tau$). In this steady-state condition (Fig.~\ref{fig-3}(e)), all unit cells have time-varying fields with identical phase differences from each other and a relatively constant maximum amplitude, satisfying the condition expressed in (2). However, during the initial time at approximately 50\,ns (Fig.~\ref{fig-3}(d)), some of the unit cells have a small amplitude because of the maximization of the absorption during the initial time. This nonlinear time-varying absorption is, however, influenced by the resonance condition of the unit cell. As we can see from the results for the initial time (Fig.~\ref{fig-3}(d)), the lowest amplitude is observed when the unit cell is tuned to near the resonance frequency, i.e., $L_{add} = $9.7\,nH, 12\,nH or 13.5\,nH. However, further from the resonance frequency, the absorption level is minimal; hence, the voltages appear the same for both an incident pulse and an incident CW, as seen, for example, with $L_{add} = $1\,nH and 40\,nH. This indicates that tuning the resonance by varying $L_{add}$ affects the time response of the unit cell. In particular, extreme resonance tuning causes the unit cell to always have the maximum amplitude level irrespective of time, as in the case of $L_{add} = $1\,nH and 40\,nH. Such unit cells with non-time-varying amplitudes are important in enabling the metasurface to behave as a diffraction grating under transient conditions, as the metasurface may otherwise undesirably absorb all incident pulses.

% === III. RIS Performance Evaluation ========================
% ==============================
\section{RIS Performance Evaluation}
To demonstrate the versatility of our method, we conduct an EM simulation of the RIS using ANSYS Electronics Desktop 2020 R2 to validate the pulse-width-based beamforming concept. As shown later in this section, we utilize the EM simulation results to model the wireless link channel of the RIS and perform a corresponding bit-error rate (BER) analysis in a communication system based on binary phase shift keying (BPSK) modulation.

\subsection{Surface Impedance Modeling of the RIS}
To implement beamforming, we consider the RIS to have an $x$-axis size of 800\,mm ($\sim$8$\lambda$), equivalent to 5 periods of the supercell (40 unit cells), and a $y$-axis size of 480\,mm ($\sim$4.8$\lambda$), equivalent to a uniform arrangement of 16 identical unit cells. This configuration yields an RIS that anomalously reflects an incident wave over the $xz$ plane (see Fig.~4(a)), with 640 unit cells in total. To simplify the EM simulation while still capturing the physical mechanism involved with the operation of the metasurface, we evaluate the RIS by using the time-dependent surface impedance model \cite{fathnan2022method}. This simulation method is equivalent to retrieving the RIS's response for different discrete pulse widths.

\begin{figure}[t]
  \begin{center}
  \includegraphics[width=\linewidth]{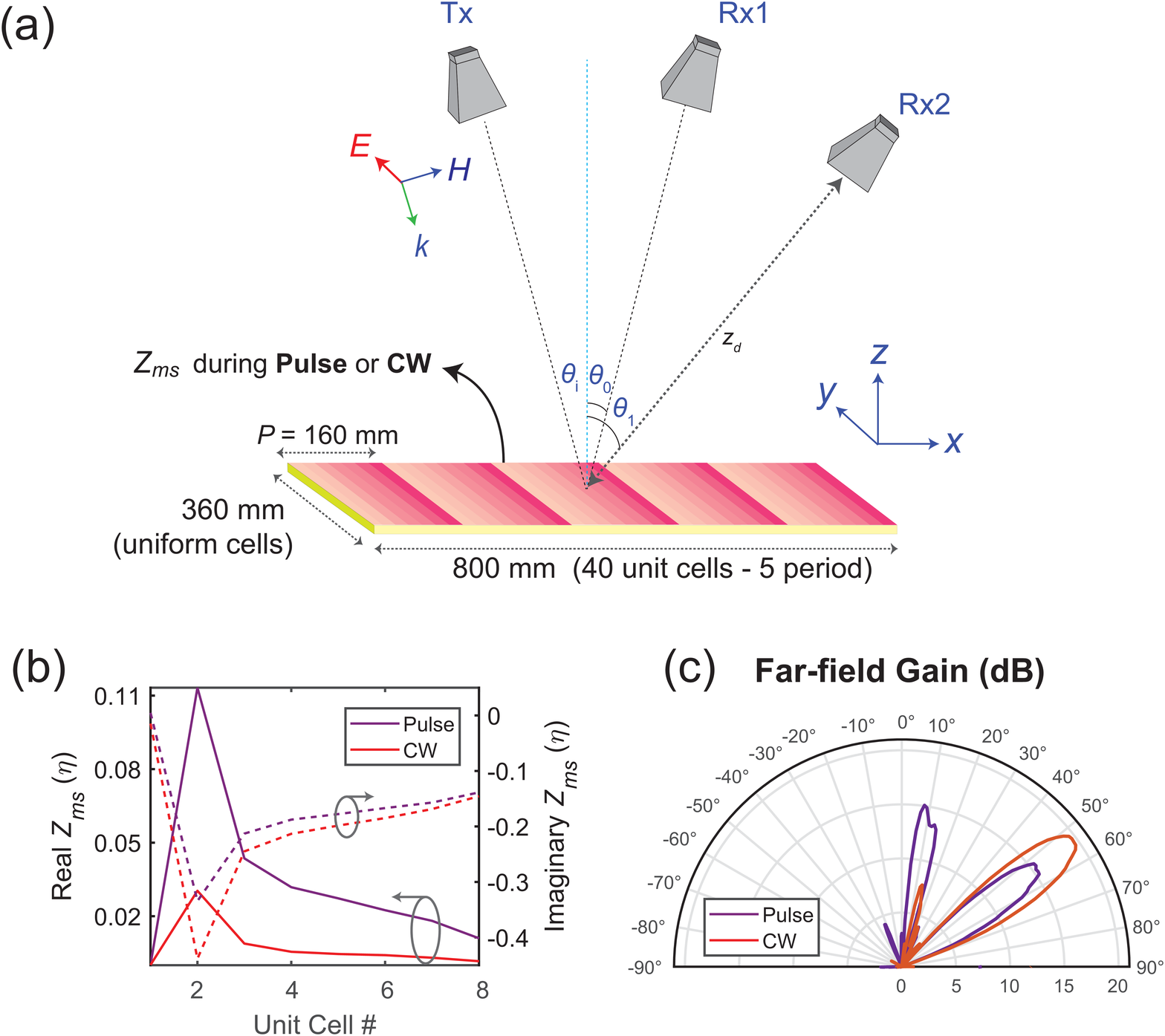}
  \caption{{(a) Simulation setup for simulating the RIS using its equivalent surface impedance. When a realistic communication scenario is considered (as in the simulation method used for Fig.~5), three horn antennas are included (Tx, Rx1, and Rx2). Unit cell numbering starts from left to right of the \textit{x}-axis. (b) Real part (solid) and imaginary part (dashed) of the surface impedance extracted from the RIS unit cells. (c) Simulated far-field gain of the RIS considering different incident waveform types (pulse and CW).}}
  \end{center}
\end{figure}

While the surface impedance model for a waveform-selective metasurface discussed in \cite{fathnan2022method} was for a transmissive case, a similar method can also be used for a reflective metasurface. For a reflective metasurface, the surface impedance can be extracted from the unit cell reflection coefficient as follows \cite{fathnan2020bandwidth,diaz2017generalized}
\begin{equation}
\begin{aligned}
{Z_{{ms}}}&=\frac{jZ_{\mathrm{11}}Z_s\tan{(\beta_s d)}}{jZ_s\tan{(\beta_s d)}-Z_{\mathrm{11}}},
\end{aligned}
\label{eq:zms}
\end{equation}
where $Z_s$ is the impedance of the dielectric substrate, $\beta_s$ is the substrate wavenumber, $d$ is the dielectric thickness, and $Z_\mathrm{11}$ is the $Z-$parameter for the reflection coefficient $S_{11}$. Therefore, using the reflection coefficient $S_{11}$ obtained from the unit cell simulation (Fig.~3(d)), we can extract $Z_{ms}$ under the corresponding incident waveform condition, either a pulse or a CW. The retrieved $Z_{ms}$ results for the proposed metasurface are presented in Fig. 4(b), where it is clearly seen that the real part of $Z_{ms}$ in the pulse case is larger than that in the CW case. This is consistent with the time-dependent reflection amplitude of the unit cell, which manifests as higher losses during the initial time.

Following the above procedure, we simulate the RIS using ANSYS HFSS to obtain the far-field profiles under the two incident waveform conditions (pulse and CW). Here, the RIS is simulated as a surface impedance on top of a grounded dielectric layer with a thickness of $d$. The real and imaginary values of the impedance shown in Fig. 4(b) are used to represent the corresponding unit cells. The surrounding material of the RIS is specified as a vacuum region with open boundaries. A plane wave source is set at the front face of the vacuum box with an incidence angle of $\theta_i=-10^\circ$ facing the RIS. The simulation results are presented in Fig.~4(c), where we see that the far-field profiles are consistent with the aforementioned beamforming technique with pulse-width-based design. Note that in this simulation  the horn antennas shown in Fig.~4(a) are not used. In the case of an incident pulse, the metasurface reflects at approximately the same power level (15\,dB gain) at both the zeroth and first diffraction orders. In the CW case, the metasurface is optimized for reflection at the first diffraction order ($\theta_1 = 53^\circ$), with a gain of approximately 20\,dB, while reflection at the zeroth diffraction order ($\theta_1 = 10^\circ$) is suppressed by 13\,dBm (i.e., 7\,dB gain).

\subsection{RIS System Model Using BPSK Modulation}
To further evaluate the communication performance of the RIS, we simulate the RIS in a realistic wireless environment. The whole system model involves the BPSK modulation scheme with a structure that consists of a single transmitter at the input and two receivers at the output of the wireless link. Within the wireless link, the RIS performs beamforming to modify the signal intensity at the two receivers.

Fig.~5(a) shows the block diagram of the system model, where $b(t)$ denotes the input binary data and $\omega_c$ is the angular frequency of the carrier signal (3\,GHz). Each bit is phase modulated by a cosine function with a signal length of $T_b = 5$\,ns and a binary phase condition ($\pm{90^\circ}$). On the output side, each received signal is summed with an additive white Gaussian noise (AWGN) component $n(t)$, followed by multiplication with the carrier signal and integration to remove the harmonics. The demodulation process ends with a threshold detector reconstructing the received signal into the corresponding bits.

\begin{figure}
  \begin{center}
  \includegraphics[width=\linewidth]{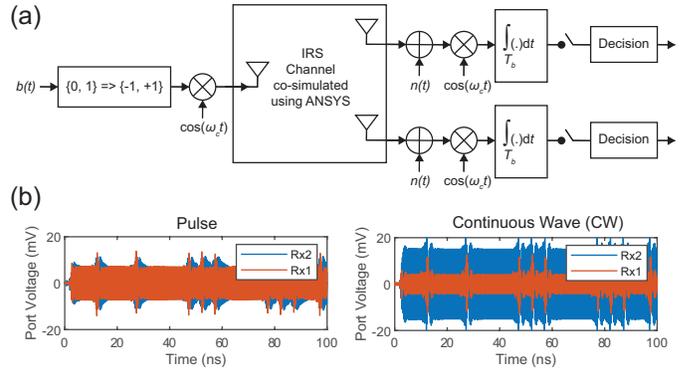}
  \caption{{(a) Block diagram of the RIS-based communication system using BPSK modulation. (b) Simulated BPSK signals received at Rx1 and Rx2 from 0 to 100\,ns.}}
  \end{center}
\end{figure}

An EM model is used to represent the wireless link channel in order to evaluate the pulse-width-dependent beamforming operation of the RIS. Here, following the method outlined in Section~III-A, a more realistic communication configuration using three horn antennas is simulated (see Fig.~4(a)). A transmitting horn antenna (Tx) is placed at a distance of $z_d = 1.6$\,m from the RIS and tilted by $-10^\circ$ in the $xz$ plane. Two other horn antennas, representing two different receivers (Rx1 and Rx2), are placed at the same distance $z_d$ from the metasurface, one tilted by $\theta_0 = 10^\circ$ and the other by $\theta_1 = 53^\circ$ in the $xz$ plane, corresponding to the zeroth and the first diffraction orders of the RIS. Each horn antenna is equipped with a wave port, and these wave ports serve as the input (Tx) and two outputs (Rx1 and Rx2) for the corresponding BPSK modulation. To incorporate the BPSK-modulated signal into the simulation, first, we perform co-simulation with ANSYS Circuit and ANSYS HFSS. Then, the generated BPSK signals from the transmitter block (obtained from MATLAB) are used at the Tx port.

The signals acquired from the simulation (Rx1 and Rx2) are then processed using MATLAB for demodulation. Because we consider AWGN at the receiver ends, it is possible to conduct BER analysis based on the variation in the signal-to-noise ratio (SNR). The results are shown in Fig.~6, where we see that the BERs at the two receiving antennas vary depending on the incident waveform. In the pulse case, the BER at Rx1 is slightly lower than that at Rx2, indicating that both antennas receive a broadcast signal from Tx. However, in the CW case, the BER at Rx2 is substantially smaller than that at Rx1, indicating that the wireless signals are mainly directed toward Rx2. This result confirms that the RIS controls the wireless channel conditions depending on the pulse width to support either broadcast or unicast communication, without the need for complicated control lines within the RIS. 

\begin{figure}
  \begin{center}
  \includegraphics[width=0.8\linewidth]{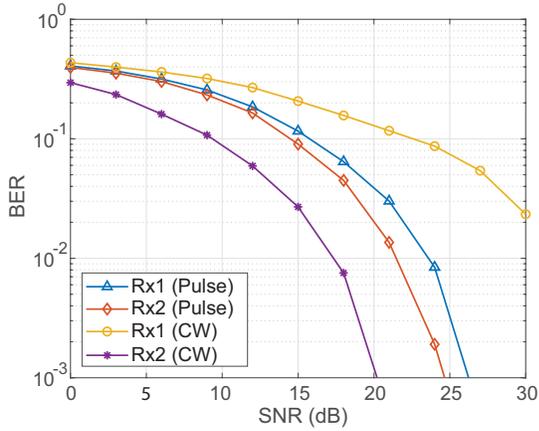}
  \caption{{BER vs. SNR analysis using the proposed RIS-controlled wireless channel. The two receiver antennas have different BER profiles for different waveform types, validating the proposed beamforming technique with pulse-width-based design.}}
  \end{center}
\end{figure}

\section{Conclusion}
A method of controlling the beamforming operation of an RIS is proposed based on the pulse width variation of the incident wireless signals. The RIS configuration here represents a prospective scheme for addressing the synchronization problems faced in the existing RIS-based systems with only simple hardware requirements and a simple control algorithm. The proposed RIS is based on a waveform-selective metasurface consisting of patches embedded with nonlinear and transient circuits on top of a grounded dielectric layer. The RIS reflects an incident CW with the maximum amplitude and a gradient phase profile spanning 2$\pi$ in one period. For short pulses, the RIS reflects with the maximum amplitude from only some unit cells, giving rise to a sparsely radiating condition that results in two diffracted waves with nearly identical power. This self-tuning mechanism allows the RIS to produce different beam patterns for different incident waveforms. To verify this scheme, an RIS system model using BPSK modulation is considered, in which the RIS-controlled wireless link is connected to a single transmitter at the input and two receivers at the output. Numerical demonstrations involving an EM model of the RIS-controlled wireless channel successfully validate the concept of beamforming based on the incident pulse width. More specifically, the beam pattern of the proposed RIS is controlled by tailoring the reflection in a passive manner that depends on the pulse width, which does not require any control lines connected to the unit cells or precise synchronization between the BS and the RIS.

\ifCLASSOPTIONcaptionsoff
  \newpage
\fi

\bibliographystyle{IEEEtran}
\bibliography{IEEEabrv,Bibliography}

\vfill

\end{document}